\documentclass{article}

\usepackage{geometry}
\usepackage{multicol,graphicx,amssymb,mathrsfs,amsmath,pifont,amscd,latexsym,color,fancyhdr,CJK,url,longtable, pifont, subfigure,epstopdf,setspace,multirow,colortbl,tabularx,threeparttable, booktabs, verbatim, bbm,listings, balance,color,indentfirst}
\usepackage[linesnumbered,boxed,algosection]{algorithm2e}

\usepackage[linesnumbered,boxed,algosection]{algorithm2e}
\newcommand{\tabincell}[2]{\begin{tabular}{@{}#1@{}}#2\end{tabular}}

\begin{document}
\lstset{
basicstyle=\ttfamily,
columns=fullflexible,
showstringspaces=false,
keywordstyle= \color{ blue!70},commentstyle=\color{red!50!green!50!blue!50},
frame=shadowbox,
rulesepcolor= \color{ red!20!green!20!blue!20}
}

\title{Roaming across the Castle Tunnels: an Empirical Study of Inter-App Navigation Behaviors of Android Users}

\author{Ziniu Hu, Yun Ma, Qiaozhu Mei, Jian Tang}

\maketitle

\begin{abstract}

Mobile applications (a.k.a., apps), which facilitate a large variety of tasks on mobile devices, have become indispensable in our everyday lives. Accomplishing a task may require the user to navigate among various apps. Unlike Web pages that are inherently interconnected through hyperlinks, mobile apps are usually isolated building blocks, and the lack of direct links between apps has largely compromised the efficiency of task completion. In this paper, we present the first in-depth empirical study of inter-app navigation behaviors of smartphone users based on a comprehensive dataset collected through a sizable user study over three months. We propose a model to distinguish informational pages and transitional pages, based on which a large number of inter-app navigation are identified. We reveal that developing ``\textit{tunnels}'' between of ``isolated'' apps has a huge potential to reduce the cost of navigation.  Our analysis provides various practical implications on how to improve app-navigation experiences from both the operating system's perspective and the developer's perspective. 


\end{abstract}

\section{Introduction}\label{sec:introduction}
Alice was wandering through restaurant reviews in the Yelp app. It mentioned that the restaurant reminded people of a classical scene in the movie ``\textit{Pretty Woman},'' which made her very eager to watch the clip. All she had to do was to go back to the OS home screen and then launch the Youtube app. When the landing page of the app was loaded, she looked for the search bar, typed in a query, and navigated through a few results before finding the clip of the scene. And the moment was gone.

Mobile applications (a.k.a., apps) are already indispensable in our everyday lives. It is reported that the traffic from mobile devices has already surpassed that from PCs and apps have become the major entrance to the Internet~\cite{MobileComparison}. Mobile users usually need to ``navigate'' among a set of apps to complete a specific task~\cite{Suman:Mobicom16}. For example,  one may receive a piece of news in the email app, read it in the newsreader app, and share it to the social networking app. Such a process is quite similar to browsing through Web pages. However, compared to the Web users who can easily navigate through the \textit{hyperlinks}, app users like Alice often have to go through a frustrating procedure to manually switch from one app to another.  
This frustration is amplified when there are more and more apps installed on a device and when the user needs to switch back and forth between various apps.

The inter-app navigation is indeed non-trivial for user interaction and has been drawing a lot of attention. 
Some solutions have been proposed to help bridge the ``isolated''  apps~\cite{Parate2013Practical,Yan2012Fast,Baeza2015Predicting}.
In particular, the recent concept of ``deep links'' has been proposed to facilitate the navigation from one app to another. Essentially, deep links are the URLs that point to specific locations inside an app page, which launches the app if it has been already installed on the device. Today, all major mobile platforms, including Android, iOS, and Windows, have supported deep links, and have been encouraging developers to implement and define the deep links to their apps~\cite{BingAppLinking,FacebookAppLinks,GoogleAppIndexing,deeplink}. So far, the deep link has been reported to have a poor coverage~\cite{Azim2016uLink} - only approximately 25\% of apps provide deep links and only a small number of pages within an app, as predefined by the developers, can be directly accessed via deep links.

Deep link is a desirable concept, but what is holding it back? While there are historical reasons such as the fear of ``stolen page views'',  the most straightforward reason is that it takes non-trivial manual effort to do so. Indeed, unlike hyperlinks which are standardized and facilitated by the HTTP, there is no gold standard for deep links. The app developers, the platforms, and the apps stores all have to spend substantial effort to define, implement, standardize, and maintain them. Without teleportation spell, tunneling through the castles means tedious human work.


Is it that bad? Perhaps not really. In practice, navigation between apps is actually performed between the ``\textit{pages}'' inside the apps. Analogical to a website, an app also contains various pages, e.g., the landing page, advertisement pages, content pages, etc. Not all apps need to be deeply linked, and for those that do, not every single page needs to be linked from outside the apps. 
If one can distinguish those ``important'' pages from the rest and understand their relationship in navigation, the effort can be significantly reduced. 
Developers may focus their effort on creating deep links for these pages; they may also target the ``partner'' apps/pages. 
Furthermore, if the next page that a user is likely to visit can be predicted and a method that can  dynamically generate deep links is available~\cite{Azim2016uLink}, runtime facilities such as Avitate~\cite{Baeza2015Predicting} and FALCON~\cite{Yan2012Fast} can be leveraged to navigate users to the destination page more efficiently. 
At the minimum, even if either of the wishes comes true, a simple estimation of the potential benefit of deep links could make the ``deep link advocacy'' more persuasive. Answers to these questions exactly requires an in-depth analysis of the inter-app navigation behavior, which unfortunately does not exist in literature.

This paper makes the first empirical study of the page-level inter-app navigation behavior of Android users. We present an in-depth analysis of inter-app navigation based on a three-month behavioral data set collected from 64 users in 389 Android apps, consisting of about 0.89 million records of app-page level navigation. Inspired by the Web search and browsing, we classify the \textit {informational} pages where users tend to stay and interact for long at an app, and the \textit{transitional} pages where users only stay awhile or tend to skip when they navigate among informational pages. We summarize various patterns of inter-app navigation which can imply some routines of users.
We believe that our empirical study results can provide useful insights to various stakeholders in the app-centric ecosystem.

The main contributions of this paper are:

\begin{itemize}
\item We propose a classification model to distinguish pages in an app into informational pages and transitional pages. We study the distribution of users' staying time of these two kinds of pages. The staying time in transitional pages looks like a \textit{Gaussian} distribution with a mean of 4.3 seconds and a small value of variance, while the staying time of informational pages follows a \textit{log-normal} distribution with the mean staying time of 29.4 seconds and a large variance, indicating that the time spent on this type of pages can differ dramatically.

\item We then demonstrate the inefficiency caused by transitional pages in inter-app navigation. The average time cost of inter-app navigation is around 13 seconds when navigating among different apps, in which the transitional pages account for 28.2\%. Such an overhead of inter-app navigation is non-trivial for mobile users.
 \item We explore two frequent patterns in inter-app navigations by analyzing which apps/pages are more likely to be linked during inter-app navigation, and under what contexts such navigation would happen. The task-specific pattern indicates that apps can be classified into clusters under which the apps cooperate with each other to accomplish specific tasks. The contextual pattern indicates that the inter-app navigation is related to context information such as network type, time, location, etc.
 \item We propose some practical implications based on the findings, to facilitate app developers, OS vendors, and end-users. For example, we make a proof-of-concept demonstration by employing a machine learning based approach to accurately predicting the next informational page from current state and thus recommend a navigation path between two pages (i.e., a potential deep link) to reduce unnecessary transitional pages.
\end{itemize}

To the best of our knowledge, this is the first empirical study on inter-app navigation behavior at a fine-grained level, i.e., page level rather than app level. The rest of this paper is organized as follows. Section~\ref{sec:motivation} presents the background of Android apps and formulate the inter-app navigation at the page level. Section~\ref{sec:data} presents our behavioral data set and how it was collected. Section~\ref{sec:activity} describes an empirical study of inter-app navigation at page level and characterizes the navigation patterns. Section~\ref{sec:implication} discusses some practical and potential useful implications that can be explored based on our empirical study. Section~\ref{sec:discussion} discusses the limitations of our work. Section~\ref{sec:related} relates our work to existing literature and Section~\ref{sec:conclusion} concludes the paper.

\section{Inter-App Navigation: In a Nutshell}\label{sec:motivation}

An Android app~\cite{AndroidGuide}, identified by its \texttt{package name}, usually consists of multiple \texttt{activities} that are related to each other. An \texttt{activity} is a component that provides an interface for users to interact with, such as dialing phones, watching videos, reading news, or viewing maps. Each activity has a unique \texttt{class name} and is assigned a window to draw its graphical user interface. 


\begin{table}[!t]
\centering\small
\caption{Conceptual comparison between Android apps and Web}\label{table:concepts}
\begin{tabular}[ht]{|c|c|c|}
 \hline
  \tabincell{c}{\textbf{Concepts of Android Apps}} & \tabincell{c}{\textbf{Concepts of Web}} & \textbf{Example}\\
  \hline
  app & website & youtube\\ \hline
  package name & domain & www.youtube.com\\ \hline
  activity & Web page template & \tabincell{c}{https://www.youtube.com/watch?v=[xxx]}\\ \hline
  activity instance & Web page instance & \tabincell{c}{https://www.youtube.com/watch?v=qv6UVOQ0F44}\\ \hline
\end{tabular}
\end{table}

For ease of understanding, we can draw an analogy between Android apps and the Web, as compared in Table~\ref{table:concepts}. An Android app can be regarded as a website where the package name of the app is similar to the domain of the website. An activity can be regarded as a template of Web pages and an instance of an activity is like a Web page instance. Different Web pages of the same template differ in the values of parameters in their URLs. For example, URLs of different video pages on the Youtube website have the format of \url{https://www.youtube.com/watch?v=[xxx]}. This template can be regarded as the VideoActivity in the Youtube app. Without loss of generality, in this paper, we use the term \emph{page} or \emph{app page} to represent the activity that has a UI for users to interact with on their smartphones.



In order to accomplish a task, users usually have to navigate through many pages. Among them, different pages serve different roles. Inspired by studies on the Web~\cite{broder2002taxonomy, LiuTSC09}, we classify the pages of mobile apps into two categories:

\begin{itemize}
    \item \textbf{informational page:} These pages serve to provide content/information, such as news article pages, video pages, mail editing page or chatting page, etc. Users accomplish their desired intention of one app in its informational pages.

    \item \textbf{transitional page:} These pages are the intermediate pages along the way to reach the informational pages. Some of the pages serve to narrow the search space of possible informational pages and direct the users to them, including the system transitional pages, such as Launcher, where users can select the desired app that may contain the potential informational pages, and in-app transitional pages, such as ListPage in a news app, where users can choose a topic to filter the recommended articles. Other pages are somehow less helpful for navigation from the users' perspective, such as the advertisement pages and the transition splash pages.
\end{itemize}

Based on our definition, the app navigation is the process of reaching the target informational page from the source informational page, via multiple transitioanl pages. Specifically, inter-app navigation is the navigation whose source and target informational pages belong to different apps.

\begin{figure}[t!]
\centering
  \includegraphics[width=0.9\textwidth]{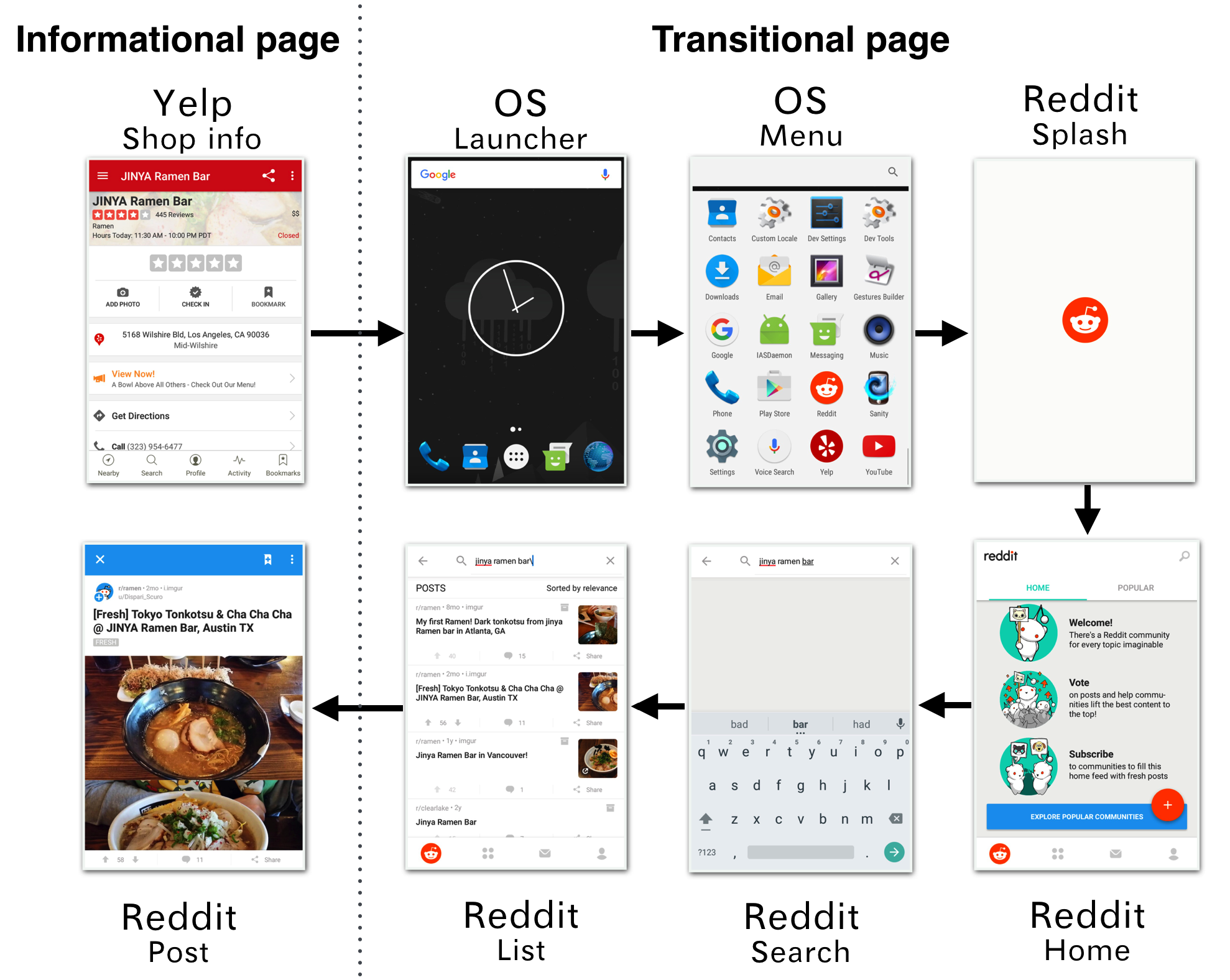}
  \caption{An example showing a user's navigation process from a ramen restaurant page in Yelp to read comments in Reddit.}~\label{fig:snip}
\end{figure}

To illustrate the navigation process, a simple scenario is drawn in Fig~\ref{fig:snip} . Suppose a user has found a ramen restaurant in the Yelp, and wants to read detailed comments about this restaurant in Reddit. To accomplish this task, the user has to first quit Yelp app to the OS launcher, click the app menu and then opens the Reddit app. Afterwards, the user has to start from the home page of Reddit app, walking through a series of transitional pages such as Search page and List page, and finally reaches the post page to read the comment about this restaurant. In this scenario, the true intention of the user is to be directly navigated to the target informational page, i.e., the post page, to read the comment. However, the user needs various transitional pages before landing on the final target page.

Intuitively, users would like to spend more time on informational pages and avoid transitional pages. In view of that, some popular apps adopt flat UI design and carefully organize its functionalities and information display in order to reduce the number of transitional pages for reaching the target informational page. However, due to the fact that most of the apps provide only dedicated functionalities, users usually need to switch among multiple apps to achieve one task, resulting in notable time cost on inter-app navigations.

We formally define a page-level app navigation as a triple $<s, t, \psi>$, where $s$ is the source informational page, $t$ is the target informational page, and $\psi=\{p_i\}$ is a sequence of transitional pages that represent a navigation path from $s$ to $t$. We ensure that the screen state of the smartphone is always \emph{On} in a whole navigation process. $\psi$ can be empty, indicating that there is a direct link from $s$ to $t$ without having to pass through any transitional pages. By the definition, the inter-app navigation is a special type of app navigation where $s$ and $t$ belong to different apps.

\section{Data Collection}\label{sec:data}

To study the inter-app navigation on Android smartphones, we conduct a field study by collecting behavioral data from real-world users. In this section, we present the design of our data collection tool and the description of the data set.

\begin{figure}[t!]
\centering
  \includegraphics[width=0.6\textwidth]{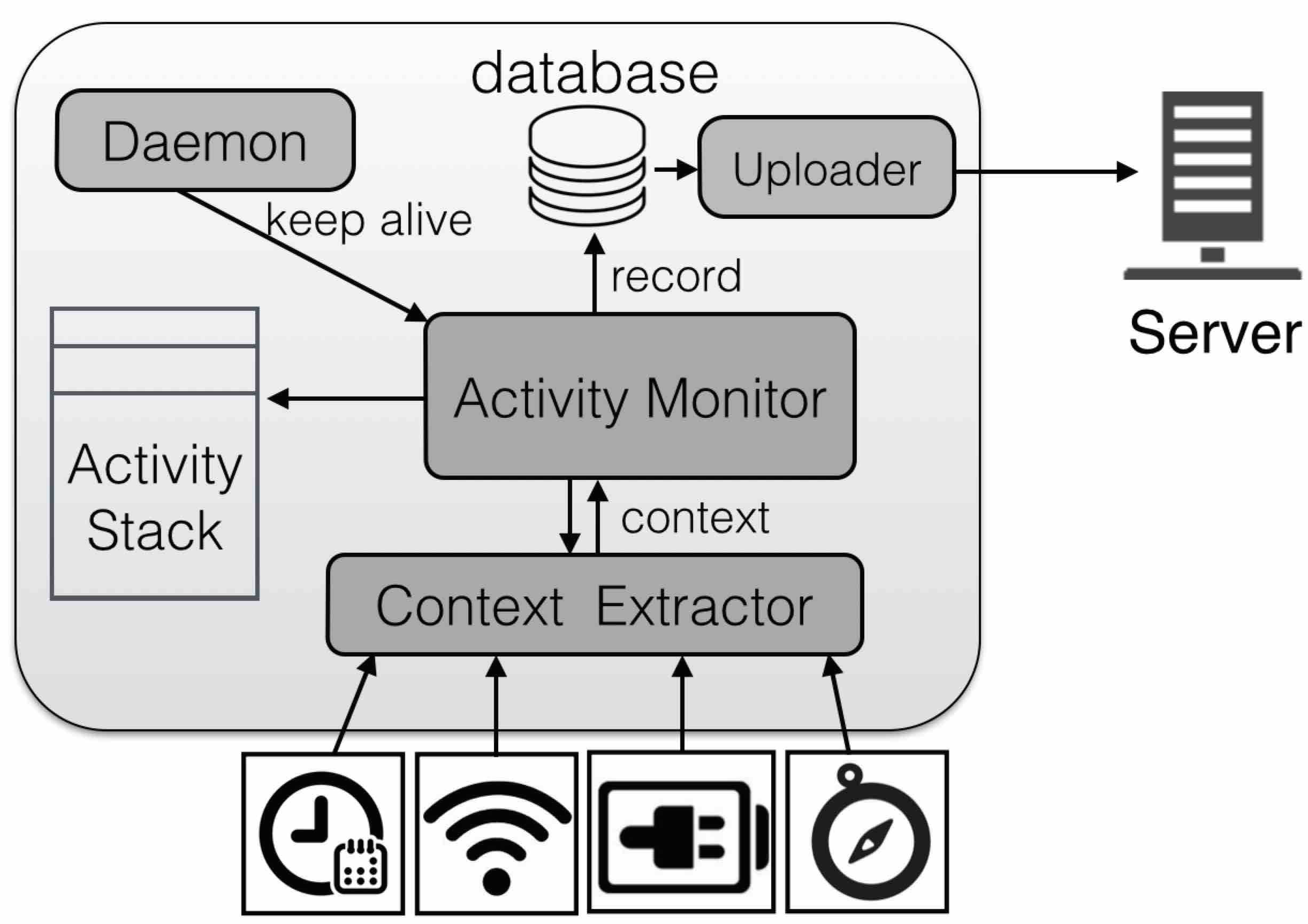}
  \caption{The architecture of the data collection tool}~\label{fig:tool}
\end{figure}

\begin{table*}[t!]
\footnotesize
\caption{Data example}
\label{tab:data_table}
\centering
\begin{tabular}{llllllll}
    \hline
    \textbf{User ID} & \textbf{App} & \textbf{Activity} & \textbf{Time} & \textbf{Network} & \textbf{Screen} \\
    \hline
    99000478917655 & com.eg.android.AlipayGphone & fund.ui.FundMainNewActivity & Mar. 31, 15:25:0 & cellular & ON \\
    \hline
    353925062095766 & com.tencent.mm & ui.LauncherUI & May 2, 22:17:13 & Wi-Fi & ON \\
    \hline
    863473022828516 & com.android.contacts & activities.ContactDetailActivity & May 9, 14:49:26 & Off & OFF \\
    \hline
    357523051693018 & com.sina.weibo & page.NewCardListActivity & Feb. 11, 15:20:50 & cellular & ON \\
\hline
\end{tabular}
\end{table*}

As mentioned above, we focus on the page-level inter-app navigation. To this end, we develop a tool to monitor the system events representing such behaviors, as shown in Fig~\ref{fig:tool}. The tool is a monitoring app running at the background of Android platforms. It consists of four modules, \emph{Activity Monitor}, \emph{Context Extractor}, \emph{Daemon}, and \emph{Uploader}. The activity monitor reads the top of the activity stack from the system every one second and produces a record entry whenever the top activity changes, indicating that a transition between app pages occurs. In the meantime, the monitor invokes the context extractor to collect the user's context information, including network type (cellular/Wi-Fi/Off), local time and screen status (ON/OFF). The records are stored in a local database, and the uploader will upload the records to our server once a day at night and under Wi-Fi network condition. Therefore, the tool does not influence the normal usage of mobile devices. To keep our tool from being killed by the system, the daemon module periodically checks the status of the tool and re-launches it if necessary. To protect user privacy, we anonymize the device ID with a hash string.

We recruited student volunteers for the data collection via an internal social network site in Peking University. We got 64 on-campus student volunteers who fully agreed with the collection statements, and we installed the tool on their Android smartphones. The data collection lasted for three months, and we finally collected 894,542 records, containing 3,527 activities from 389 apps\footnote{The data collection and analysis  process was conducted with IRB approval from the Research Ethic Committee of Institute of Software, Peking University. We plan to release the collected data once the work is published.}.


Table ~\ref{tab:data_table} provides some illustrating examples:  \texttt{User ID} denotes the unique and anonymized identifier of the user; \texttt{App} and \texttt{Activity} denote which app and page (i.e., activity) that the user is interacting with, respectively. \texttt{Time} refer to the local Beijing time when the page is visited. \texttt{Network} indicates the network type when the page is visited, i.e., cellular, Wi-Fi, or offline; \texttt{Screen} denotes the status of the device screen, i.e., ON and OFF.

\section{Empirical Analysis}\label{sec:activity}
According to the definition of inter-app navigation, the most important issue is to identify the informational pages from transitional pages. In this section, we present an empirical study on our collected user behavioral data, based on which we can identify the informational pages and transitional pages through a clustering approach. We then explore the characteristics and find some interesting patterns of inter-app navigations.




\subsection{Clustering the App Pages}

According to the definition in Section~\ref{sec:motivation}, to identify the inter-app navigation, we should distinguish informational pages from transitional pages in our collected records of page usage. Just like the similar experiences on the Web pages~\cite{DBLP:conf/sigir/LiuGLZMHL08}, a simple but intuitive measure is based on the length of staying time spent on the pages. Intuitively, users are likely to spend longer time on the informational pages, while the transitional pages are only for navigation purpose and thus users are likely to spend rather short time on them. Since our data records the sequence of pages, we are able to calculate the time interval for every single page by the timestamp when this page is visited and the timestamp when its subsequent page is visited. As each page could appear several times, we can obtain the distribution of the staying time for each page.

Table~\ref{tab:toplist} lists example pages with the longest and shortest average staying time. From the name of these pages, we can speculate that the pages with longer staying time are more likely to be the pages that provide substantial information and available services. While the pages with shorter staying time have names containing ``list'' or ``launcher'', indicating that these pages are more likely to be transitional pages. These examples imply that the staying time can be a vital clue to identify transitional pages.
\begin{table*}[t!]
\centering
\footnotesize
\caption{Top five pages with shortest and longest average staying time}
\label{tab:toplist}
\begin{tabular}{l|lllc}

\hline \textbf{Category}  & \textbf{Page Name}               & \textbf{App Name}   & \textbf{Page Description} & \textbf{Average st} \\ \hline
\multirow{5}{*}{\emph{Shortest Staying Time}}
& lenovo.Launcher                  & Lenovo System       & System Home Screen          & 1.08s               \\

& kugou.LockScreenActivity         & Kugou Music         & Music LockScreen          & 1.28s               \\
& dazhihui.InitScreen              & dazhihui Investment & App Home Page             & 1.67s               \\
& miui.home.Launcher               & Miui System         & System Home Screen          & 1.72s               \\
& hupu.ListNewActivity & Hupu Sports         & News List Overview        & 1.85s               \\ \hline
\multirow{5}{*}{\emph{Longest Staying Time}}
& mobileqq.TextPreviewActivity        & QQ               & Chatting Interface     & 73.6s \\
& android.AlarmAlertFullScreen        & System           & System Clock           & 72.1s \\
& papd.HealthDailyPopActivity         & Pingan Doctor    & Healthy Daily News     & 66.3s \\
& renren.ChatContentActivity   & Renren           & Chatting Interface     & 64.4s \\
& meitu.mtxx.MainActivity             & Meitu Xiu Xiu    & Photo Beautification   & 62.8s \\ \hline
\end{tabular}
\end{table*}

Inspired by the work by Van \textit{et al.}~\cite{Van2016A}, which propose a classification approach to determine the seesion time threshold of user interaction, we propose to use an unsupervised clustering approach to separate the pages according to the users' staying time on the pages. A simple approach may be employing the average staying time by all users as the threshold. However, simply using the average staying time cannot reflect the overall staying time distribution of mobile users. Thus, we represent each page with the probability distribution of all visits' staying time.

\begin{figure}[t!]
\centering
  \includegraphics[width=0.75\textwidth]{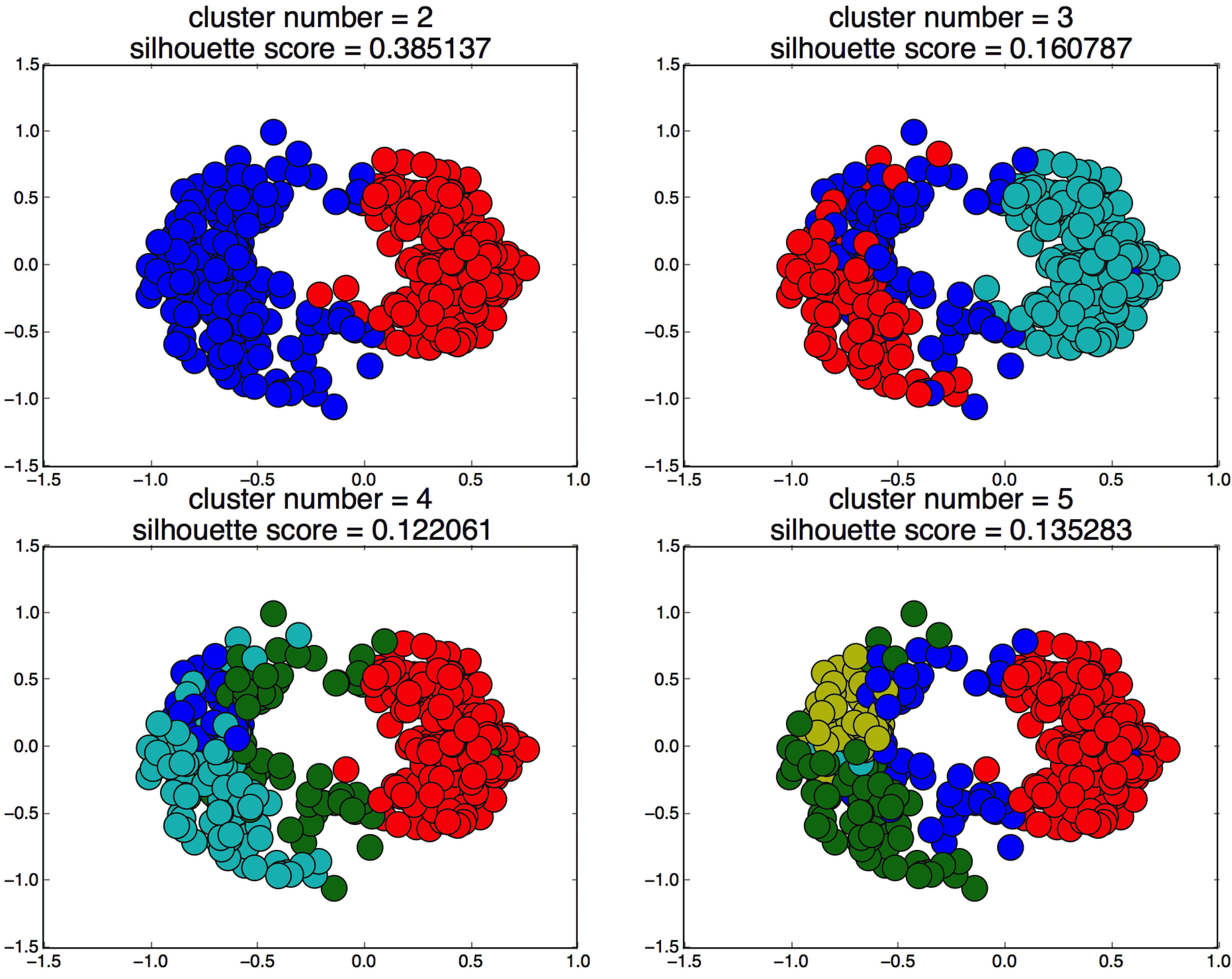}
  \caption{Clusters of pages. Visualized using the Fruchterman-Reingold force-directed algorithm. The optimal cluster number is 2.}~\label{fig:cluster_img}
\end{figure}

To measure the distance between the staying time distribution of pages, we may use the classic Kullback-Leibleer divergence (KLD)~\cite{Kullback1987The} and its symmetric version, the Jensen-Shannon divergence (JSD)~\cite{Fuglede2004Jensen}. In this paper, we choose the JSD distance as the distance between the distribution of every pairs of pages because its symmetric property makes it more appropriate for cluster. Then, we use the spectral clustering method~\cite{Ng2002On} to cluster the pages.  More specifically, such a process first learns a low-dimensional representation of each page according to the distance matrix between the pages, and then deploys the K-means algorithm to cluster the pages based on the low-dimensional representations. We assign different numbers of clusters in the K-means algorithm and select the optimal number of clusters through the Silhouette score~\cite{Rousseeuw1987Silhouettes}. Fig.~\ref{fig:cluster_img} visualizes the clustering results with different numbers of clusters using Fruchterman-Reingold force-directed algorithm~\cite{Kobourov2012Spring}. The best clustering results are obtained when the number of clusters equals to 2, which well verifies our intuition that the pages can be classified into two categories.



\begin{figure}[t!]
\centering
  \includegraphics[width=0.5\textwidth]{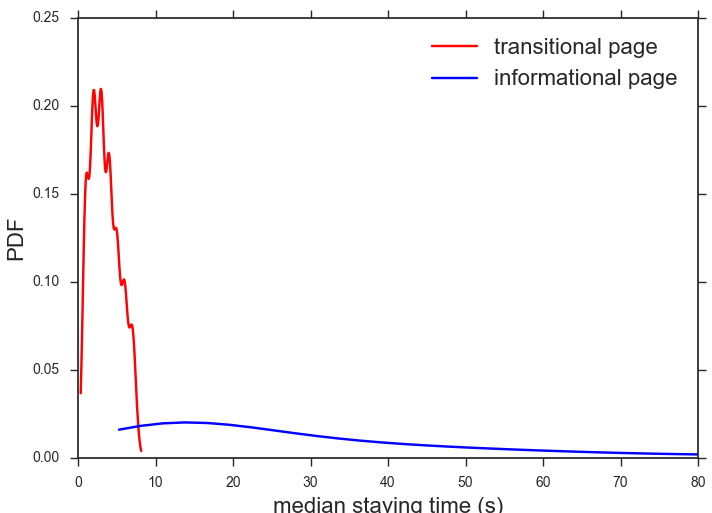}
  \caption{Probability density distribution of staying time of two types of pages.}~\label{fig:cluster}
\end{figure}

To further validate the distribution of staying time of these two types of pages, for each page, we calculate the median of its staying time and then depict the Probability Density Function (PDF) of all the pages in Fig~\ref{fig:cluster}. The blue curve indicates the PDF of staying time for informational pages while the red curve indicates transitional pages. The distribution of the transitional pages's staying time looks like a Gaussian distribution with a mean of 4.3 seconds and a small value of variance. The distribution of informational pages' staying time follows a log-normal distribution with the mean staying time of 29.4 seconds and a large variance, indicating that the time spent on this type of pages can differ dramatically.

To evaluate the authenticity of our cluster, we randomly select 100 activities, and manually classify these activities by their name, staying time and screenshots. It turns out that this result is completely accorded with our cluster labels, verifying the reliability of our result.

After identifying informational and transitional pages, we extract all the navigations from the records. Since we focus on only the navigation between different apps, we select all the inter-app navigation. The amount of inter-app navigation records is 41,619, which accounts for 26.9\% of the total navigations. Our following analysis is conducted on these inter-app navigation records.

\subsection{Characterizing Inter-App Navigation}

Ideally, users prefer to be directly navigated between two informational pages, without any transitional pages on the navigation path. 
To explore whether the current inter-app navigation performs in such user-desired fashion, we use two metrics to quantitatively measure: 1) \emph{time cost}, which is the aggregated value of staying time for all the transitional pages between the source informational page and target informational page; 2) \emph{step cost}, which is the number of transitional pages on a navigation path.

\begin{figure}[t!]
\centering
\subfigure[CDF curve of step cost.]{
    \label{fig:deep:pdf1} 
    \includegraphics[width=0.42\textwidth]{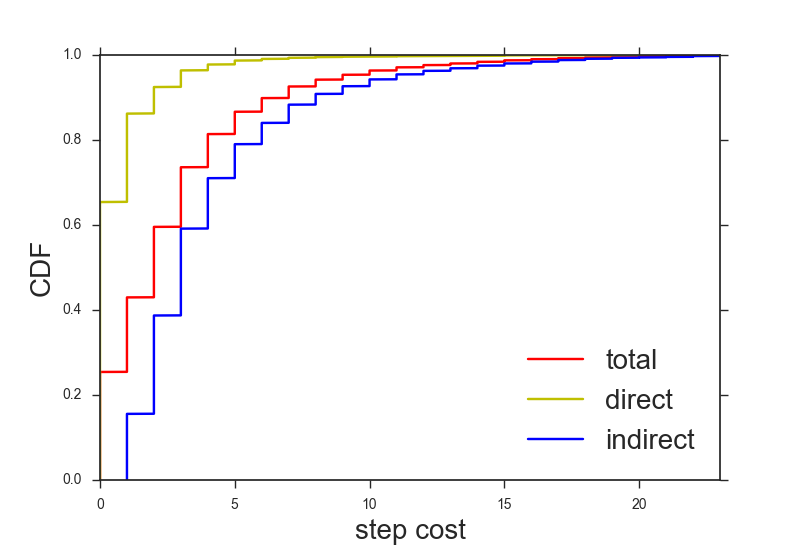}
}
\subfigure[CDF curve of time cost.]{
    \label{fig:deep:pdf2} 
    \includegraphics[width=0.42\textwidth]{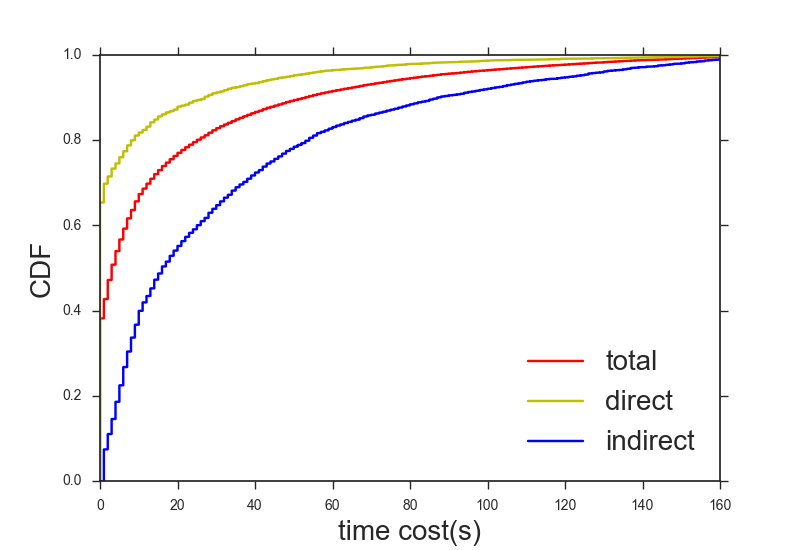}
}
\caption{Overhead of inter-app navigation}
~\label{fig:deep} 
\end{figure}

The distribution of \emph{time cost} and \emph{step cost} for all the inter-app navigations are illustrated as the red curve and labeled as \textbf{total} in Fig~\ref{fig:deep}. The average \emph{time cost} is 13.01s, and the average \emph{step cost} is 2.69, indicating that  users have to transit about 3 pages and spend 13 seconds when navigating among different apps. Additionally, we find that the time cost on the transitional pages takes up to 28.2\% of the total time in the entire inter-app navigation, which is the total time cost of the source informational page, the subsequent transitional pages and the target informational page during a single navigation process. Therefore, \textbf{ the overhead of inter-app navigation is non-trivial for mobile users.}

We observe that most of the inter-app navigations do not jump directly from the source app to the target app. Instead, a large number of inter-app navigations involve pages from the Android system such as \texttt{android.Launcher}, which is the system home screen, or \texttt{android.RecentsActivity}, which is the system list page showing all the recently used apps. We call these navigations as indirect navigations. However, a small number of inter-app navigations do not contain these system pages, but consist of transitional pages only from the source and target apps, e.g., the landing page of the target app. We call this kind of navigation as direct navigation between apps.

Comparing the distribution of \emph{time cost} and \emph{step cost} between direct navigations and indirect navigations in Fig~\ref{fig:deep}, we have the following observations:

\begin{itemize}
    \item The average time cost of direct and indirect navigations are 8.97 seconds and 24.43 seconds respectively; the average step cost of direct and indirect navigation are 0.97 and 3.86 respectively. \textbf{This result indicates that the overhead of direct navigations is much smaller than that of indirect navigations}.
    \item All the indirect navigations have at least one intermediate step, i.e., the transitional page belongs to the Android system. In contrast, 66\% of direct navigation have no such cost, indicating no navigation overhead. Therefore, the direct navigation is more appreciated.
\end{itemize}

We further explore how direct navigations could happen. Some links navigate to the system-wide apps such as Call, Maps, Camera, and so on, which are pre-installed in the OS and have special APIs to invoke. Other links navigate the users directly from the source page to the target page of third-party apps without passing through system navigational pages, such as Launcher. There links are likely to be implemented by the emerging popular \textit{deep link}~\cite{deeplink}. Similar to hyperlinks of Web pages, deep links also employ URI to locate pages and can be executed to directly open the target page from the source page. For example, when a user reads an interesting article and wants to share it to facebook, he/she can click a share button, invoking a deep link that directly navigate he/she to the facebook interface without walking through multiple transitional pages.

We are interested in whether the direct navigation is achieved by means of deep link. To this end, we first label the category information of each app\footnote{In this paper, we simply use the categorization system from a leading Android market, called Wandoujia~\url{http://www.wandoujia.com}.}, and filter out the system-wide apps. The remained third-party apps accounts for 82.7\% of the total apps. Next, we check the manifest file of the third-party apps that contain the target pages to ensure they provide deep link interfaces. Finally, we employ a popular Android analysis tool IC3~\cite{bhoraskar2014brahmastra} to check whether the source page actually invokes the target page's deep link interfaces.
Given two apps, IC3 can compute all the inter-component communications (ICC) between these apps. If there exists an ICC between the activities of the source page and target page, then we can confirm that the source page has an entry to invoke the target page's deep link interfaces. Results show that within the direct navigation pairs, all the target pages provide deep link interfaces and all the source pages have an entry to invoke the corresponding interfaces, indicating that these direct navigations are highly likely to be implemented by deep links.

\subsection{Patterns of Inter-App Navigation}


Next, we explore whether there exist any patterns of inter-app navigations. By analyzing which apps/pages are more likely to be linked during inter-app navigation, and under which contexts such navigation would happen, we find two kinds of navigation patterns: \emph{task-specific pattern} that the inter-app navigation is related to a specific type of task, which is inferred by page property, and \emph{contextual pattern} that the inter-app navigation is related to context information such as network type, time, location, etc, which is the characteristic feature of mobile usage.

\subsubsection{Task-Specific Navigation Pattern}

\begin{figure}[t!]
\centering
  \includegraphics[width=0.62\textwidth]{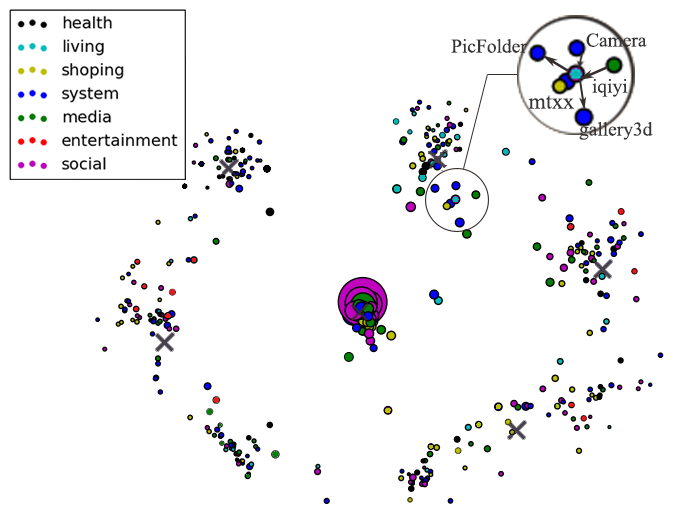}
  \caption{Inter-app navigation graph connecting informational pages from different apps. Each node is labeled by the category of the app. Edges are not displayed to clealy show patterns.}~\label{fig:points}
\end{figure}

We draw a graph representing all inter-app navigations by removing those transitional pages. As is illustrated in Fig~\ref{fig:points}, each node corresponds to an independent informational page, and is labeled with the category information of the app that the page belongs to. We assign an edge between two nodes from different apps if there exists a navigation between them. We then measure the distance by the static transition probability between every single informational page, which is calculated by the frequency of a particular navigation divided by the total navigation times. In this way, we get the distance matrix of the graph and visualize it with the classic Fruchterman-Reingold force-directed algorithm~\cite{Kobourov2012Spring}, where each node size is determined by its out-degree.

Interestingly, there are some significant clusters. In particular, there is a ``core'' cluster, in which most of the pages belong to the SOCIAL, SYSTEM, and MEDIA apps. Other clusters around the core include smaller nodes from different categories. From the distance matrix, we find that those small clusters around the core are mutually isolated with each other, while the nodes in these clusters are strongly cohesive. Additionally, all these small clusters have a strong correlation with the core cluster. We also find that apps consisting these small clusters belong to different categories.

For example, Fig~\ref{fig:points} shows the pages of one cluster that is zoomed in. We can see that these pages come from a photo beautification app called \texttt{MeiTuXiuXiu} (mtxx for short), a gallery app called \texttt{PicFolder},  a video app called \texttt{iQiyi}, system camera and gallery, along with the directed links among them. Such observations may reflect the following scenario: a user captures a screenshot from a video page of \texttt{iQiyi} or takes a photo using the camera, then he/she navigates to the edit page in the \texttt{mtxx} and decorates this image. Eventually he/she saves the image into the \texttt{PicFolder}.

Such observations indicate that the inter-app navigation can be intuitively classified into some clusters. Each cluster could represent a specific user-task scenario, where different pages may cooperate with one another to accomplish this task.

Indeed, the current results can only imply that such patterns may consist of a task. In our future work, we plan to combine session-level analysis to comprehensively specify the task characteristics~\cite{Halfaker:WWW2015}.

\begin{figure*}[t!]
  \centering

    \subfigure[Off Network]{
    \label{fig:network:off} 
    \includegraphics[width=0.28\textwidth]{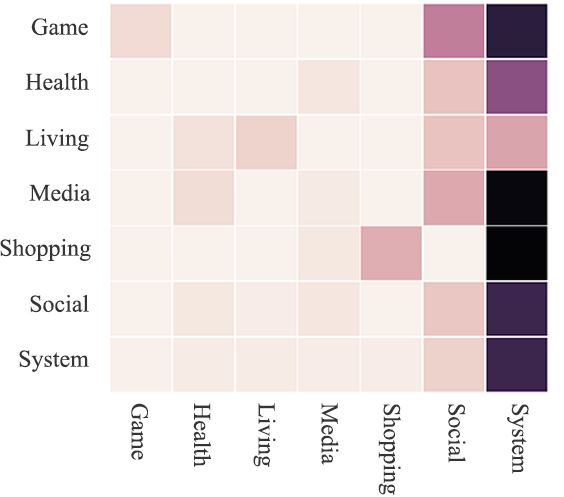}}
        \subfigure[Wi-Fi]{
    \label{fig:network:wifi} 
    \includegraphics[width=0.28\textwidth]{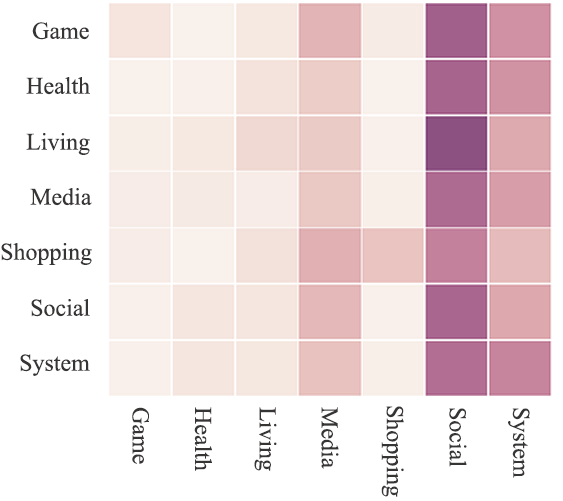}}
    \subfigure[Cellular]{
    \label{fig:network:cell} 
    \includegraphics[width=0.31\textwidth]{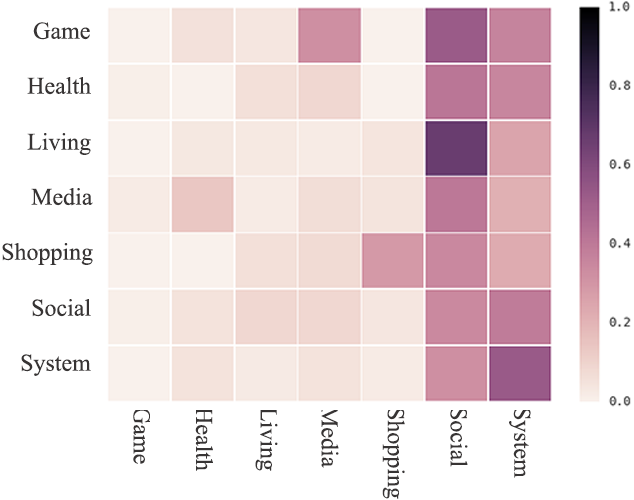}}

  \caption{Heatmaps of transitonal probability between pages from different categories. The three figures refer to the transition under different network types respectively.}
  \label{fig:network} 
\end{figure*}

\subsubsection{Contextual Navigation Pattern}

\begin{figure}[t!]
\centering
  \includegraphics[width=0.65\textwidth]{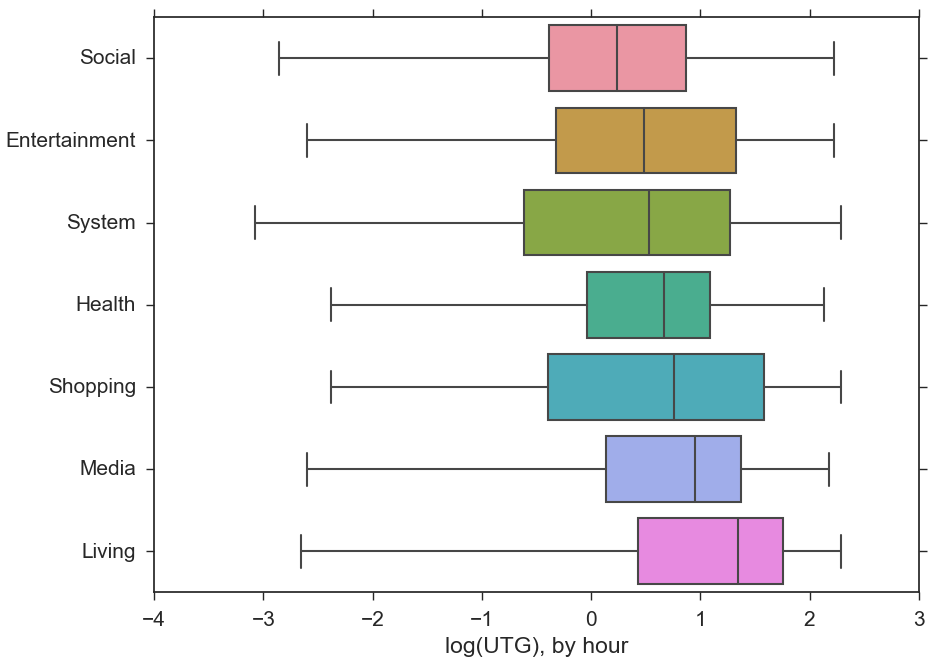}
  \caption{The boxplot of the usage time gap for seven categories.}~\label{fig:time}
\end{figure}

Next, we are interested if any contextual patterns exist, i.e., whether some navigations are more likely to occur under specific contexts. From our collected data, we focus on two types of contexts, i.e., \textit{network} and \textit{time}.

We first analyze whether the type of network affects user's preference of inter-app navigation. We classify all navigations into three clusters based on their network type. Similar to the task-specific pattern, for each cluster, we calculate the static transition probability between each pair of informational pages. Then we plot a heatmap for each network-type cluster, where pages are denoted by their categories. The source pages are plotted on the Y-axis while target pages are plotted on the X-axis. The results are presented in Fig~\ref{fig:network}.  It is observed that only the \texttt{SYSTEM} is significant when the users are offline, while all the other categories are rather shallow. It indicates that users may mainly switch between \texttt{SYSTEM} apps such as SMS, Dialing, Camera, etc., when they are offline. When the network is under Wi-Fi and cellular, the inter-app navigations are more likely to happen, and \texttt{SOCIAL} apps are usually involved. However, the frequent patterns can vary between these two network types. When users are under Wi-Fi, \texttt{MEDIA} apps are more likely to be linked with other apps. Since \texttt{MEDIA} apps such as video players may require high-bandwidth and stable connection, it makes sense that such a navigation pattern exists. In a sense, one can infer which pages/apps that are more likely to be visited given the network type.

Then we investigate whether some navigations are \textit{time} sensitive. We compute the visit frequency of a page during a specific period of time, with a metric namely Usage Time Gap (UTG), which is defined as the interval between current time when a page is visited and the time when this page was latest visited. We draw the boxplot of the UTG for each categories in Fig~\ref{fig:time}. The UTG varies among different categories. For \texttt{SOCIAL} apps, the median value and variance of UTG is the smallest, indicating that users may frequently visit the pages of these apps in a short time interval. For the \texttt{Media} and \texttt{Living} apps, their median UTG are much larger than the others, indicating that these apps are not ``daily routines" for users. In other words, when used once, these apps are not likely to be revisited by users in a short time interval. Intuitively, with the metric such as UTG, one can infer which pages are more likely to be visited. Combined with the sequential information between two informational pages, it would be possible to prefetch these pages according to the patterns captured with UTG.

\section{Implications}\label{sec:implication}

The preceding results have illustrated that the inter-app navigation has some patterns. Thus, the most intuitive implication is whether the navigation could be predicted, so that the transitional pages can be reduced or even be eliminated. Based on the prediction, our study would be useful for stakeholders including end-users, app developers, and OS vendors.

\subsection{Predicting Page Navigation}

We first investigate whether the inter-app navigation could be predicted, i.e., given a source informational page $s$, which target informational page $t$ a user is likely to $s$ navigate to? In fact, we can predict such behaviors based on our previously derived patterns. To make a proof-of-concept demonstration, we use a machine learning technique to conduct the prediction. Intuitively, such a prediction problem can be treated as a \textit{classification} problem. We can treat a pair of app page navigation $(s, t)$ as an instance. Each pair of navigation $(s, t)$ can be represented with different types of features such as the features for source informational page $s$ and target informational page $t$, as well as the contextual features when the navigation occurs.


With the actual inter-app navigation behaviors in our collected data, we can observe many positive instances of inter-app navigation pairs. However, given a source informational page $s$, it is  unknown which target information pages $t$ will never be navigated to. Theoretically, this is known as a typical one-class classification problem~\cite{Tax2001One} in literature. Here we adopt a start-of-the-art algorithm \emph{PU} proposed by Liu et al.~\cite{Liu2003Building}~\cite{Li2003Learning} to solve the problem. The basic idea of the algorithm is to use the positive data to identify a set of informative negative samples from the unlabelled data (Here, each possible pair of source and target informational pages can be treated as an unlabeled instance.) In our experiments, we compare different variants of the algorithms, including S-EM, Spy+SVM, Spy+SVM-I, NB+EM, NB+SVM, and NB+SVM-I, and a naive solution which simply samples some random target pages for each source page and treats them as the negative samples.

The overall classification process can be simply described in Algorithm~\ref{tab:algo}. In the beginning, the algorithm generates an unlabelled data set $U$ by randomly sampling some target pages $t$ for each source page $s$. Then the algorithm identifies informative negative samples according to the positive training data. For prediction, given a source page $s$, we can rank the candidate target page $t$ according to the probability of the pair $(s,t)$ belonging to the positive class.




\begin{algorithm}
    \caption{One-class classification algorithm for prediction. }
    \label{tab:algo}
    \small
    \SetAlgoLined
    \KwIn{PU Method $\mathcal{PU}$ , Positive dataset $P$, Candidate Page Set $S$=($p_1,p_2,...,p_N$)}
    \KwOut{Recommendation List $L$}
      $U \leftarrow geneate\_unlabelled\_data(P)$\\
      $N \leftarrow extract\_negative(U, P)$ according to~\cite{Liu2003Building}\\
      $\mathcal{PU}.classifier.fit(P \cup N)$\\
      \ForEach{$tp \in S$} {
          $prob[tp] = \mathcal{PU}.classifier.predict\_probability((s,t))$\\
      }
      $L \leftarrow sort\_by\_probability(P, prob)$\\
    \Return{$L$}

\end{algorithm}


\begin{table}[t!]
\small
\centering
\caption{Feature Table.}
\label{tab:feature}
\begin{tabular}{l|l|l}
\hline
\textbf{Feature Type}                  & \textbf{Feature Name}        & \textbf{Dimensions} \\ \hline
\multirow{3}{*}{\emph{task-specific feature}} & Page name           & 816*2=1632 \\
                               & app name            & 245*2=490  \\
                               & category id     & 20*2=40    \\  \hline
\multirow{2}{*}{\emph{contextual feature}}       & Network type        & 3         \\
                               & Usage Time Gap (UTG) & 1 \\\hline
\end{tabular}
\end{table}

In Table ~\ref{tab:feature}, we illustrate two types of features to represent each navigation pair $(s, t)$,  including the task-specific features of the pages and the contextual features associated with the pairs. For the task-specific features, three different features are identified including the name of the pages, the name of the apps which the pages belong to, and the name of the category that the page belongs to. The contextual features include the network type and $UTG$, representing the elapsed time of the target app since it was used last time until now. Note that all the features except $UTG$ are quantitative variables, so we process them in one-hot encoding.

We report the prediction results with different combinations of these features to illustrate their importance. As a ranking problem, we evaluate the prediction result according to a well adopted measure for ranking, called mean reciprocal rank (MRR) metric~\cite{MRR}. The MRR of a recommendation list is the multiplicative inverse of the rank for the correct answer. Obviously, the MRR score and the recommendation performance is positively correlated. Indeed, there are different variants of the PU algorithm. Therefore, we first randomly filter 30\% of the data reserved for subsequent evaluation, and then conduct a five-fold cross validation to compare different variants of the algorithms. The results show that S-EM algorithm yields the better results.

We compare the proposed approach with two other straightforward algorithms including \textit{popularity} approach and \textit{Markov chain} based approach. \textit{Popularity} approach ranks the pages according to their usage frequency, and  \textit{Markov chain} based approach is ranked by the static transition probability for one page to be navigated from the given source page. To evaluate our approach, we use the 30\% data as test set which is previously reserved for evaluation and the remaining 70\% data as training set.

\begin{figure}[t!]
\centering
  \includegraphics[width=0.6\textwidth]{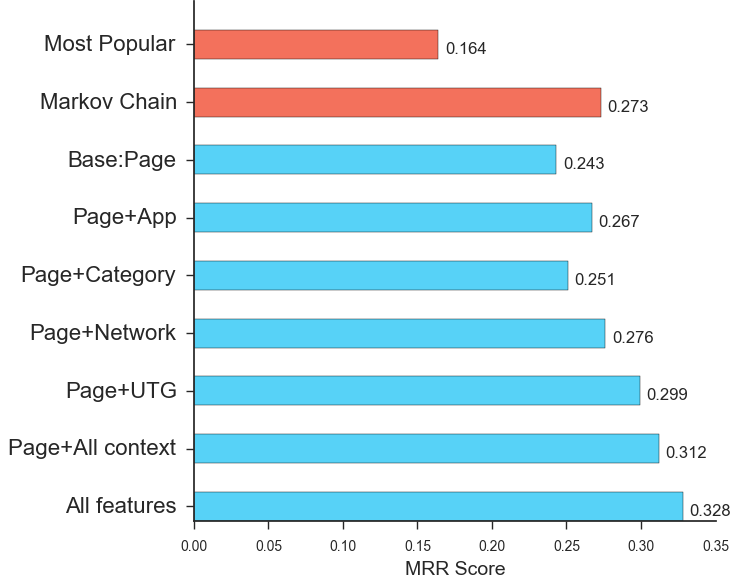}
  \caption{Prediction results with different algorithms and features.}~\label{fig:comparison2}
\end{figure}

Fig ~\ref{fig:comparison2} presents the prediction results with different algorithms and features. First, the classification algorithm with all the features outperform the naive Popularity and Markov Chain methods, with its MRR score of 0.328, implying that the actual page is on average ranked at the third place of our recommendation list. This is because the classification algorithm can effectively integrate the task-specific features of pages and contextual features for prediction, which have been proved to be very important according to the exploratory analysis in previous section.
Second, as for the importance of different features for prediction, we tried different combinations of features for prediction. We first treat the page name as the basic feature. Adding the name of the app or the category of the app only slightly improves the performance. This may be because all the three features are about the task-specifics of the pages, which are strongly correlated, and the name of the pages already well represent the task-specifics of the pages. Adding either of the contextual features significantly improve the result, showing that the contexts are indeed very important factors to affect the inter-app navigation. By combing all the features, the best-of-breed prediction result is obtained.

\subsection{Further Potential Application Scenarios}

The preceding proof-of-concept demonstration indicates that the inter-app navigation could be predicted to some extent, based on which we can further explore some potential applications for different stakeholders, including app developers, OS vendors, and end-users.

Based on the information of inter-app navigation, app developers can more efficiently identify the \textit{upstream partner} apps from which users can navigate to the current app, and  \textit{downstream partner} apps to which users can navigate from the current app. On one hand, app developers could expose ``deep links'' to upstream apps in order to attract more user visits from upstream apps. On the other hand, app developers could integrate deep links from downstream apps in order to precisely introduce users for downstream apps. Similar to the hyperlinks of Web pages, deep links can introduce more ``\textit{page visits}'' of an app. In particular, the in-app ads are the major revenue channels for app developers, hereby establishing the deep links can potentially increase the possibility of ads clicking.



For current system-wide smart assistant such as Aviate~\cite{Baeza2015Predicting}, our analysis of navigation patterns can provide insights to further improve the user experiences. Aviate predicts the next app that is likely to be used by users and displays the shortcut of apps on system home screen. However, the prediction is performed at \textit{app level}, thus users still have to manually locate the desired page by several tedious steps. Combined with our analysis, the prediction can be performed at a fine-grained level with less ``transitional'' overhead but navigate users directly to the page that they are really interested in. In addition, the predicted page can be more accurate to contexts (e.g., the current network condition), and can be prefetched into RAM to enable fast launch.

Finally, the ultimate goal of our study is to build a recommendation assistant that can predict which app pages a user desires to access, and can dynamically generate deep links for such desired pages. Thus, the user can pass through a few transitional pages in the recommendation assistant, rather than the time-consuming navigation steps. Since previous research efforts such as uLink~\cite{Azim2016uLink} have already proposed practical approaches to dynamically generating deep links for app pages, the key point of such a recommendation assistant is to accurately predict which pages should be navigated. Our prediction can provide which kind of informational pages the user is likely to visit. Based on the prediction, the assistant can recommend the possible pages and generate a deep link for each page, enabling the user to choose the desired page and navigate directly to that page.  For example, if we predict that a user is going to watch a movie, the system can offer a top recommended movie list for the user to select. Our findings can integrate the recent efforts on in-app semantic analysis~\cite{Suman:Mobicom16}, by which the system can more precisely predict which app page the user desires.



\section{Discussion}\label{sec:discussion}




This paper does make the first-step empirical study to derive preliminary knowledge of inter-app navigation. Our results come from the field study conducted on 64 student volunteers' devices for three months, with 0.89 million records of user behaviors. Indeed, such a scale is a bit  small, and the selected user group may not be comprehensive enough, e.g., reflecting the preferences of only on-campus students. In this way, the empirical results and derived knowledge may not generalize to users from other groups. However, our analysis approach and prediction techniques can still be generalized. Indeed, predicting the inter-app navigation behavior is quite meaningful in various aspects, e.g., traffic accounting and in-app ads.

Our data collection tool introduces very little additional overhead. Based on our previous industrial experiences of large-scale user study collaborated with leading app store operators~\cite{Li:WWW2016} and input method  apps~\cite{Lu:UbiComp2016}, we plan to evaluate and integrate our data collection tool in these platforms, so that we can learn more comprehensive knowledge of inter-app navigation. In addition, with the access to a large number of user profile, we can add features of user modeling to enrich the analysis of navigation patterns.

Another limitation of our empirical study is that our data collection is focused at page (activity) level. However, modern Android apps make use of fragment, which is a portion of user interfaces (e.g., tabs) in an activity and can be roughly regarded as dynamic sub-pages. Ideally, it is more comprehensive to measure the session time the user spends on each fragment and on each activity, respectively. We plan to consider conducting this measurement study in our future work.

This paper demonstrates the potential overhead caused by those transitional pages. From the user's perspective, we argue that such overhead could compromise user experiences and shall be avoided. However, from the developer's perspective, some of the transitional pages are still meaningful, e.g., containing some in-app ads to increase developers' revenues. Hence, it may not be reasonable or realistic to remove all transitional pages. In practice, we need to carefully justify whether a transitional page should be contained or not on the navigation path, e.g., allowing the developers to configure when releasing deep links, or the end users to decide by their own preferences.

\section{Related Work}\label{sec:related}
In this section, we discuss the related existing literature studies and compare with our work.
\subsection{Field Study of Smartphone Usage}
There have been some field studies to investigate user behaviors on smartphones. Ravindranath \textit{et al.}~\cite{Ravindranath2012AppInsight} developed AppInsight to automatically identify and characterize the critical paths in user transactions, and they conducted a field trial with 30 users for over 4 months to study the app performance. Rahmati \textit{et al.}~\cite{Shepard2010Livelab} designed LiveLab, a methodology to measure real-world smartphone usage and wireless networks with a reprogrammable in-device logger designed for long-term user studies. They conducted a user experiment on iPhone and analyzed how users use the network on their smartphones. In their following work~\cite{Rahmati2012Exploring}, they conducted a study involving 34 iPhone 3GS users, reporting how users with different economic background use smartphones differently. Mathur \textit{et al.}~\cite{Mathur:UbiComp2016} carried out experiments with 10 users to model user engagement on mobile devices and tested their model with smartphone usage logs from 130 users. Ferreira \textit{et al.}~\cite{Ferreira2014Contextual} collected smartphone application usage patterns from 21 participants and participants¡¯ context to study how they manage their time interacting with the device. Our field study collected user behavior data from 64 volunteers for 3 months, of which the quantity is comparable with existed studies.


\subsection{Mining Navigation Patterns}
A lot of research efforts have been made on mining the navigation patterns on the Web. Some existing literatures leveraged graph models. Borges \textit{et al.}~\cite{DBLP:conf/kdd/BorgesL99} proposed an N-gram model to exploit user navigation patterns and they used entropy as an estimator of the user sessions' statistical property. Anderson \textit{et al.}~\cite{DBLP:conf/kdd/AndersonDW02} proposed a Relational Markov Model(RMM) to model the behavior of Web users for personalizing websites. Liu \textit{et al.}~\cite{DBLP:conf/sigir/LiuGLZMHL08} proposed a method of computing page importance by user browsing graph rather than the traditional way of analyzing link graph. Chierichetti \textit{et al.}~\cite{Chierichetti:WWW2012} studied the extent to which the Markovian assumption is invalid for Web users. Other literatures use sequential pattern mining to exploit association rules. Fu \textit{et al.}~\cite{DBLP:conf/iui/FuBH00} designed a system which actively monitors and tracks a user's navigation, and applied A-priori algorithm to discover hidden patterns. Wang \textit{et al.}~\cite{Wang2004Effective} divided navigation sessions into frames based on a specific time internal, and proposed a personalized recommendation method by integrating user clustering and association-mining techniques. West \textit{et al.}~\cite{West:WWW2012} studied how Web users navigate among Wikipedia with hyperlinks. Our work is inspired by these previous efforts on the Web and we focus on how users navigate among apps on mobile system. Similar to our work, Srinivasan \textit{et al.}~\cite{Srinivasan2014MobileMiner} developed a service called MobileMiner that runs on the phone to collect user usage information. They conducted a user experiment and use a sequential mining algorithm to find user behavior patterns. Jones \textit{et al.}~\cite{Jones:2015:RAS:2750858.2807542} present a revisitation analysis of smartphone use to investigate whether smartphones induce usage habbits. They distinguish the pattern granularity into macro and micro level, and find unique usage characteristics on micro level. Both of them dig into app-level usage patterns, but our work focuses on the much deeper page-level navigation patterns rather than the app level.


\subsection{Prediction of App Usage}
Predicting the apps to be used not only facilitates mobile users to target the following apps, but also can be leveraged by mobile systems to improve the performance. Abhinav \textit{et al.}~\cite{Parate2013Practical} proposed a method similar to a text compression algorithm that regards the usage history as sequential patterns and uses the preceding usage sequence to compute the conditional probability distribution for the next app. Yan \textit{et al.}~\cite{Yan2012Fast} proposed an algorithm for predicting next app to be used based on user contexts such as location and temporal access patterns. They built an app FALCON that can pop the predicted app to home screen for fast launching. Richardo \textit{et al.}~\cite{Baeza2015Predicting} proposed Parallel Tree Augmented Naive Bayesian Network (PTAN) as the prediction model, and used a large scale of data from Aviate log dataset to train their model, and achieve high precision result.


Our work differs from previous efforts much in several aspects. On one hand, the granularity of our research is on the \emph{page} level, which is deeper than the app level. As a result, we cannot exploit the existing dataset that mainly records the user behaviors at the app level. On the other hand, we reveal that the inter-app navigation has many transitional pages, which is a tremendous noise for the purpose of predicting informational pages. Thus, we propose a clustering approach to filter out the transitional pages, and enable more precise prediction.

\section{Conclusion}\label{sec:conclusion}
\renewcommand{\baselinestretch}{0.8}
In this paper, we have presented a quantitative study of the inter-app navigation behavior of Android users. The analysis was based on behavioral data collected from a real user study, which contains nearly a million records of page transitions. We found that unnecessary transitional pages visited between two inter-app informational pages take up to 28.2\% of the time in the entire inter-app navigation, a cost that can be significantly reduced through building direct links between the informational pages. 
An in-depth analysis reveals clear clustering patterns of inter-app pages, and a machine learning algorithm effectively predicts the next informational page a user is navigating into. Our results provide actionable insights to app developers and OS vendors.   

While we have made the preliminary results of demonstrating the feasibility of predicting the next page using a standard algorithm, it is not our intent to optimize the prediction performance in this study. It is a meaningful future direction to build advanced prediction models, especially to consider the previous navigation sequences in the same session and to personalize the prediction. It is also intriguing to enlarge the scale of the user study to cover users with diverse demographics. 

\bibliographystyle{abbrv}
\bibliography{ubicomp}

\end{document}